\documentclass[aps, prd, amsmath, floats, floatfix, twocolumn, nofootinbib, showpacs]{revtex4-1}
\usepackage{graphicx}
\usepackage{color}

\newcommand{\be}{\begin{eqnarray}}
\newcommand{\ee}{\end{eqnarray}}

\begin{document}

\title{Broad K$\alpha$ iron line from accretion disks around traversable wormholes}

\author{Cosimo Bambi}
\email{bambi@fudan.edu.cn}

\affiliation{Center for Field Theory and Particle Physics \& Department of Physics, Fudan University, 200433 Shanghai, China}

\date{\today}

\begin{abstract}
It has been proposed that the supermassive black hole candidates at the 
centers of galaxies might be wormholes formed in the early Universe and 
connecting our Universe with other sister Universes. The analysis of the 
profile of the relativistic K$\alpha$ iron line is currently the only available 
approach to probe the spacetime geometry around these objects. In this 
paper, I compute the expected K$\alpha$ iron line in some wormhole 
spacetimes and I compare the results with the line produced around Kerr 
black holes. The line produced in accretion disks around non-rotating or 
very slow-rotating wormholes is relatively similar to the one expected 
around Kerr black holes with mid or high value of spin parameter and 
current observations are still marginally compatible with the possibility
that the supermassive black hole candidates in galactic nuclei are
these objects. For wormholes with spin parameter $a_* \gtrsim 0.02$, 
the associated K$\alpha$ iron line is instead quite different from the 
one produced around Kerr black holes, and their existence may already 
be excluded.  
\end{abstract}


\maketitle


\section{Introduction}

Nowadays, we have strong observational evidence that the center of every 
normal galaxy harbors a dark compact object with a mass $M \sim 10^5 - 
10^9$~$M_\odot$~\cite{bh}. While we believe that all these supermassive
objects are the Kerr black holes (BHs) predicted by General Relativity, their 
actual nature has still to be verified~\cite{rev,tests}. Dynamical measurements 
can provide robust estimates of the masses of these bodies. Combining these
results with an upper bound for their radius, it turns out that these objects are 
too heavy, compact, and old to be clusters of non-luminous bodies~\cite{maoz}. 
The non-observation of thermal radiation emitted by the putative surface 
of the BH candidate at the center of our Galaxy may also be interpreted as 
an indication that the latter has an event horizon and 
is therefore a BH~\cite{eh1} (see however Ref.~\cite{eh2}). Nevertheless, the exact 
origin of these objects is not clear: we do not understand how they could have 
become so heavy in such a short time, as we know BH candidates with a mass 
$M \sim 10^9$~$M_\odot$ at redshift $z \gtrsim 6$~\cite{super}, i.e. just about 
100~million years after the Big Bang.

A speculative possibility is that supermassive BH candidates are wormholes, 
topological connections between separated regions of the spacetime~\cite{wh}. 
They may be relics of the early Universe and may connect either two different 
regions in our Universe or two different Universes in a Multiverse model. 
Such a scenario may also explain the non-observation of thermal 
radiation, as wormholes do not have a surface. The possibility of the existence
of wormholes cannot be ruled out by general arguments, it is not in contradiction
with current observations, and the search for astrophysical wormholes could
represent a unique opportunity to investigate a multi-element Universe.

Some authors have already discussed possible ways to observationally
distinguish a Kerr BH from a wormhole~\cite{previous,th}. However, previous
work has never considered the analysis of the K$\alpha$ iron line, which 
is currently the only available technique to probe the geometry of the
spacetime around supermassive BH candidates~\cite{ka}\footnote{The 
other popular approach currently available to get information on the metric 
around BH candidates is the continuum-fitting method~\cite{cfm,disk,jet}, 
i.e. the analysis of the disk's thermal spectrum. The continuum-fitting 
method can be applied only to stellar-mass objects: the disk's 
temperature scales as $M^{-0.25}$ and for supermassive BH 
candidates the spectrum falls in the UV range, where dust 
absorption makes accurate measurements impossible.}. The K$\alpha$ 
iron line is intrinsically narrow in frequency, while the one observed appears 
broadened and skewed. The interpretation is that the line is strongly altered 
by special and general relativistic effects, which produce a characteristic 
profile first predicted in Ref.~\cite{ka1} and then observed for the first time 
in the ASCA data of the Seyfert~1 galaxy MCG-6-30-15~\cite{ka2}. In this 
paper, I will use the code discussed in~\cite{disk,iron} to compute the
profile of the K$\alpha$ iron line produced around traversable wormholes 
and seen by a distant observer. I will then compare these lines with the 
ones produced from Kerr spacetimes, to check how this technique can 
test the wormhole nature of the supermassive BH candidates in 
galactic nuclei.

\begin{figure*}
\begin{center}
\includegraphics[type=pdf,ext=.pdf,read=.pdf,width=7.5cm]{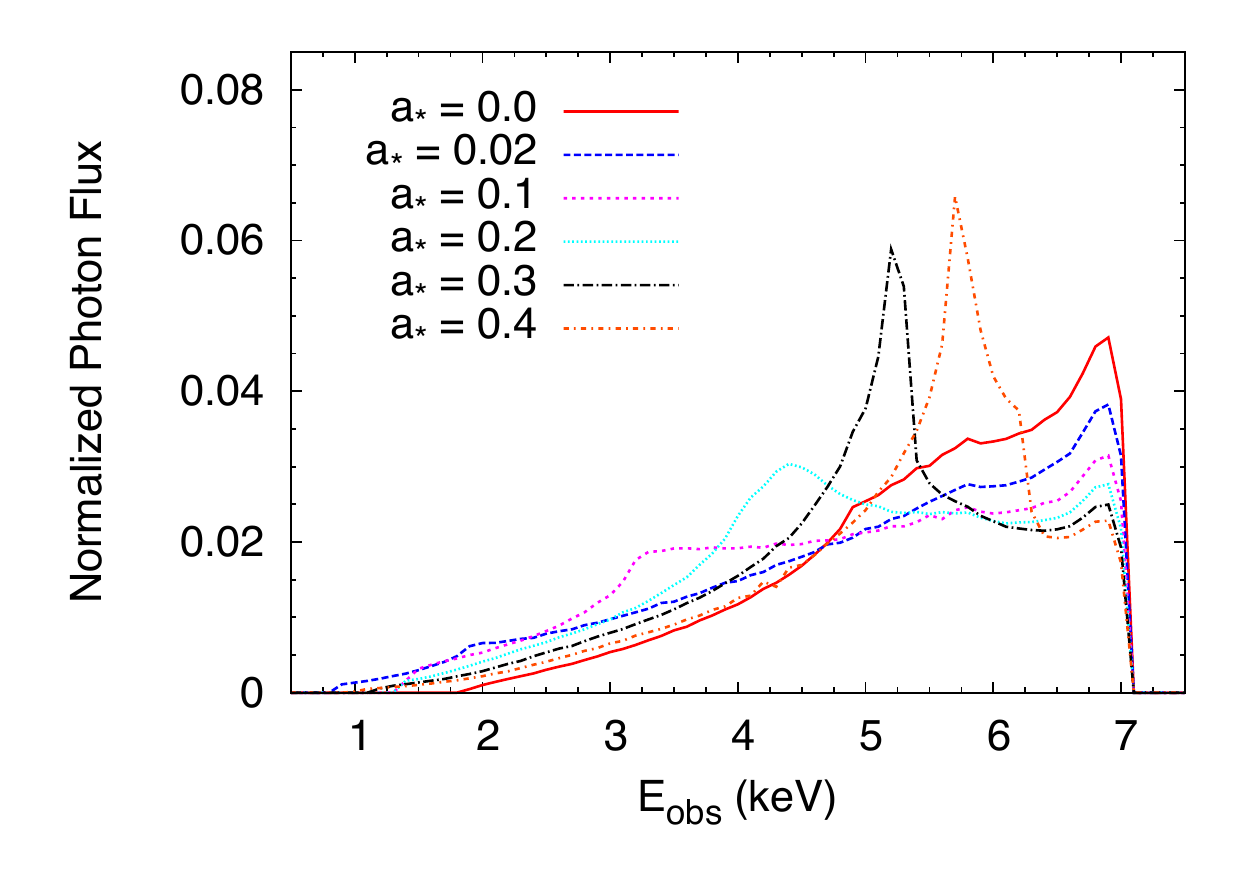}
\includegraphics[type=pdf,ext=.pdf,read=.pdf,width=7.5cm]{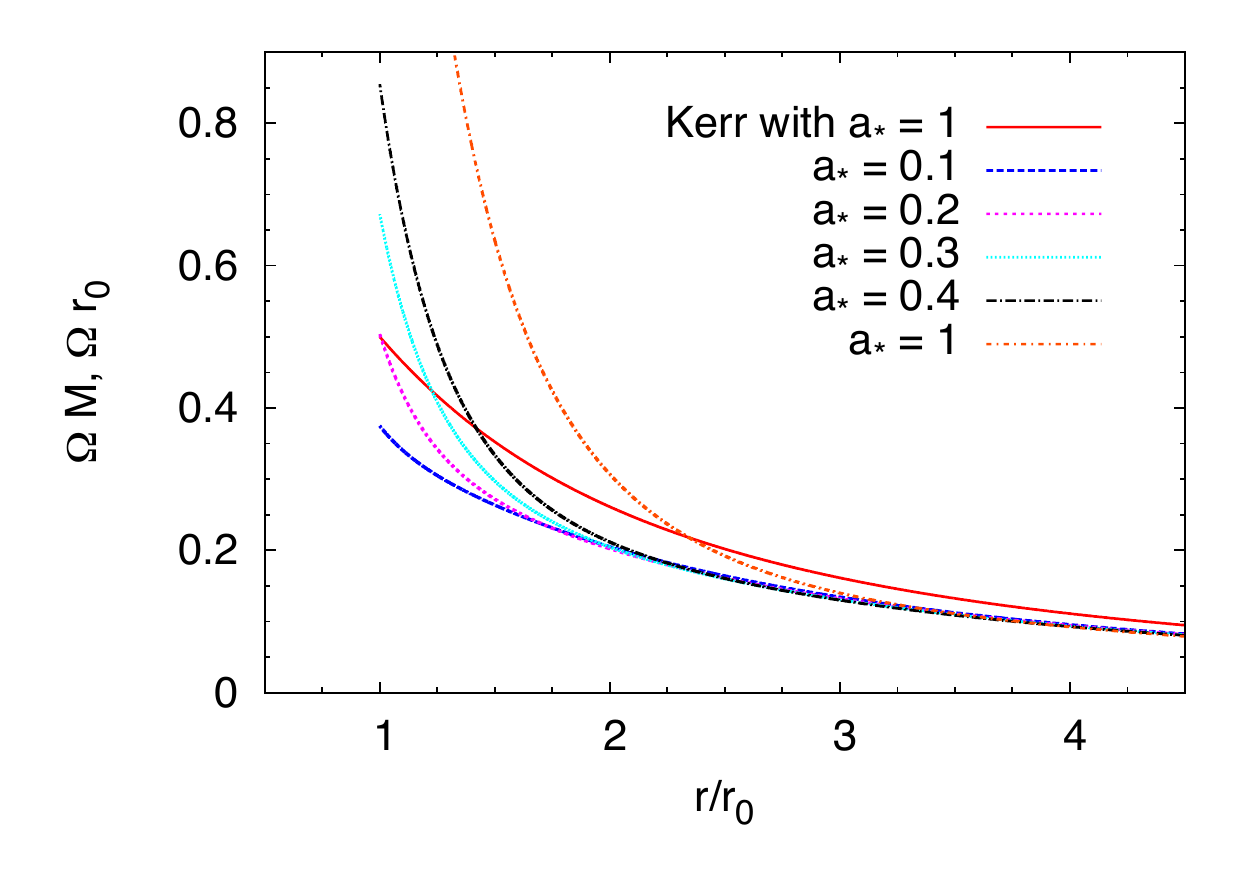}
 \end{center}
 \vspace{-0.5cm}
\caption{Left panel: broad K$\alpha$ iron line in wormhole backgrounds with 
$\gamma = 1$ and different values of the spin parameter $a_*$. The astrophysical 
parameters are: viewing angle $i = 45^\circ$, intensity profile with index 
$\alpha = -3$, and emissivity region with inner radius $r_{\rm in} = r_{\rm ISCO}$ 
and outer radius $r_{\rm out} = r_{\rm ISCO} + 100 \, r_0$. Right panel: angular 
frequency of equatorial circular orbits as a function of the radial coordinate 
$r$ for the wormhole spacetimes with $\gamma = 1$ and $a_*=0.1$, 0.2, 0.3, 0.4,
and 1, and the Kerr background with $a_* = 1$. See the text for details.}
\label{f1}
\end{figure*}

\begin{figure*}
\begin{center}
\includegraphics[type=pdf,ext=.pdf,read=.pdf,width=7.5cm]{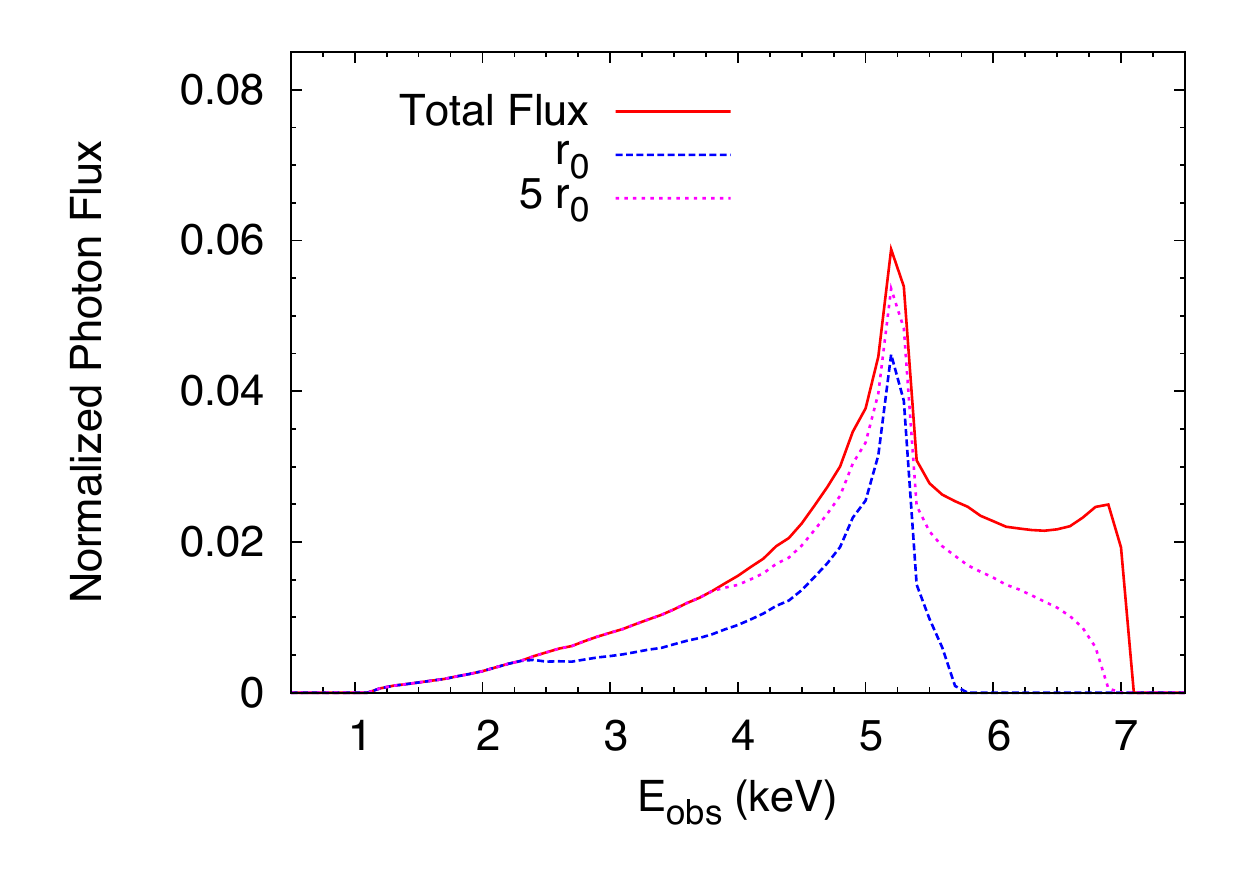}
\includegraphics[type=pdf,ext=.pdf,read=.pdf,width=7.5cm]{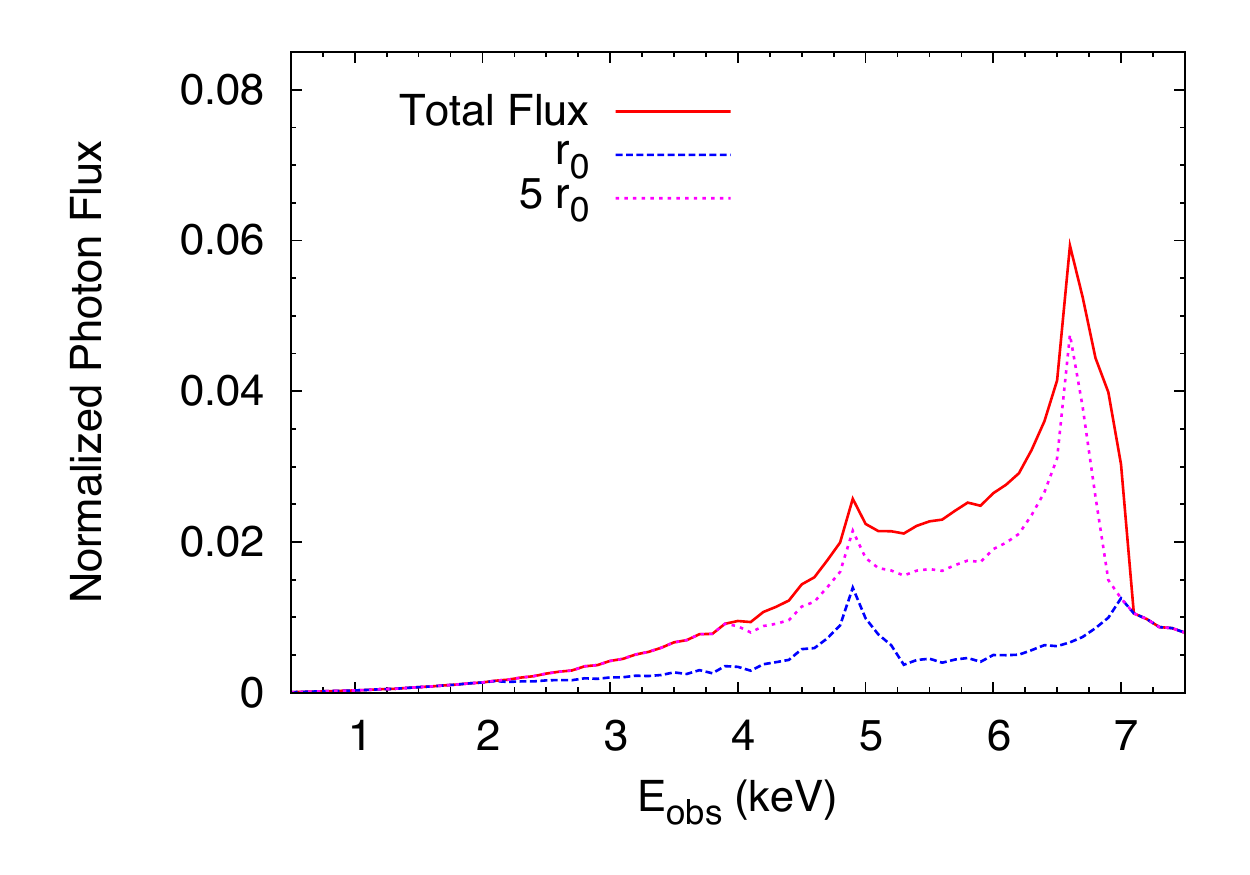}
 \end{center}
 \vspace{-0.5cm}
\caption{K$\alpha$ iron line in wormhole backgrounds with $\gamma = 1$,
$a_* = 0.3$ (left panel) and 1 (right panel), viewing angle $i = 45^\circ$, 
intensity profile with index $\alpha = -3$, and emissivity region with inner 
radius $r_{\rm in} = r_{\rm ISCO}$ and outer radius $r_{\rm out} = r_{\rm ISCO} 
+ 100 \, r_0$. The blue dashed line shows the contribution of the photon
emitted in the region of the disk from the inner radius to $r_{\rm ISCO} + r_0$.
The magenta dotted line shows the one from the inner radius to $r_{\rm ISCO} + 
5 \, r_0$.}
\label{f1b}
\end{figure*}

\begin{figure*}
\begin{center}
\includegraphics[type=pdf,ext=.pdf,read=.pdf,width=7.5cm]{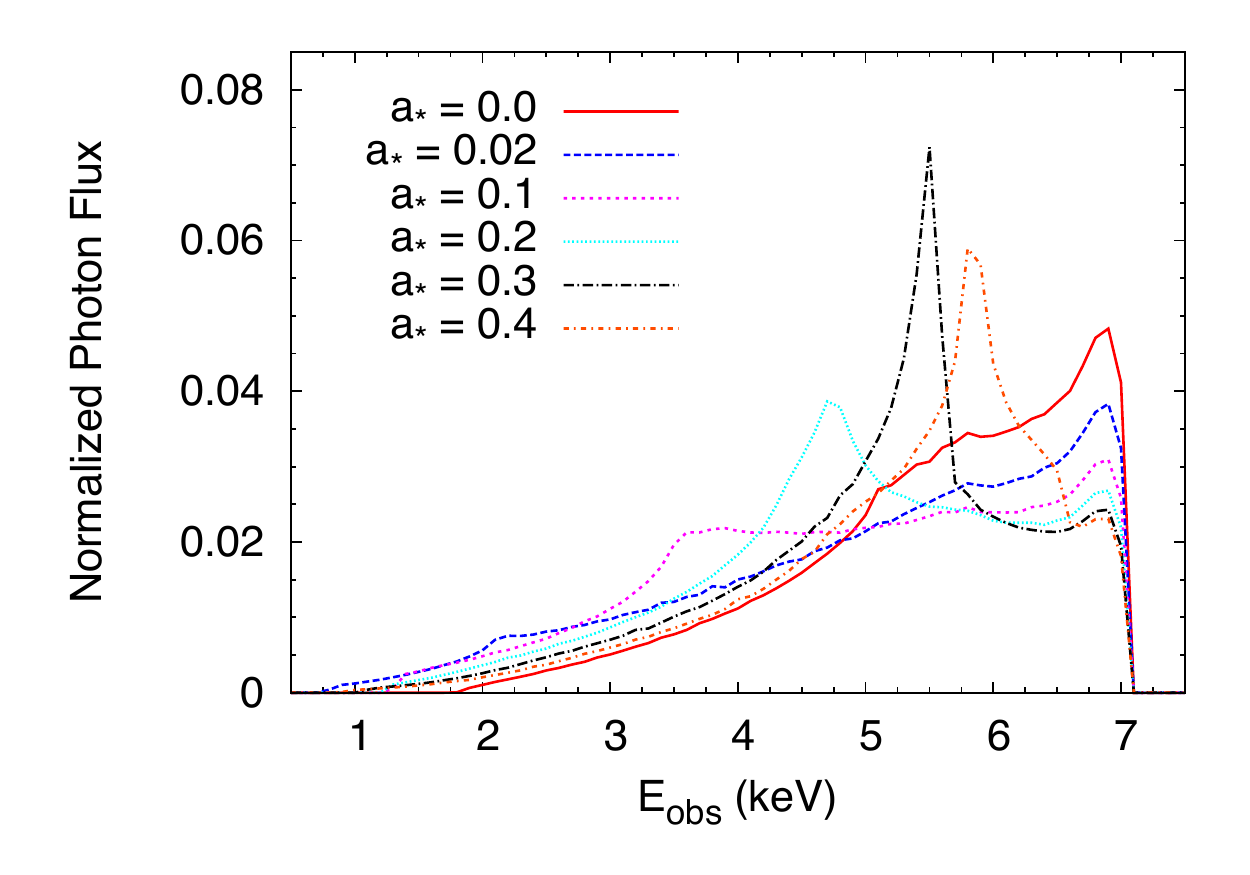}
\includegraphics[type=pdf,ext=.pdf,read=.pdf,width=7.5cm]{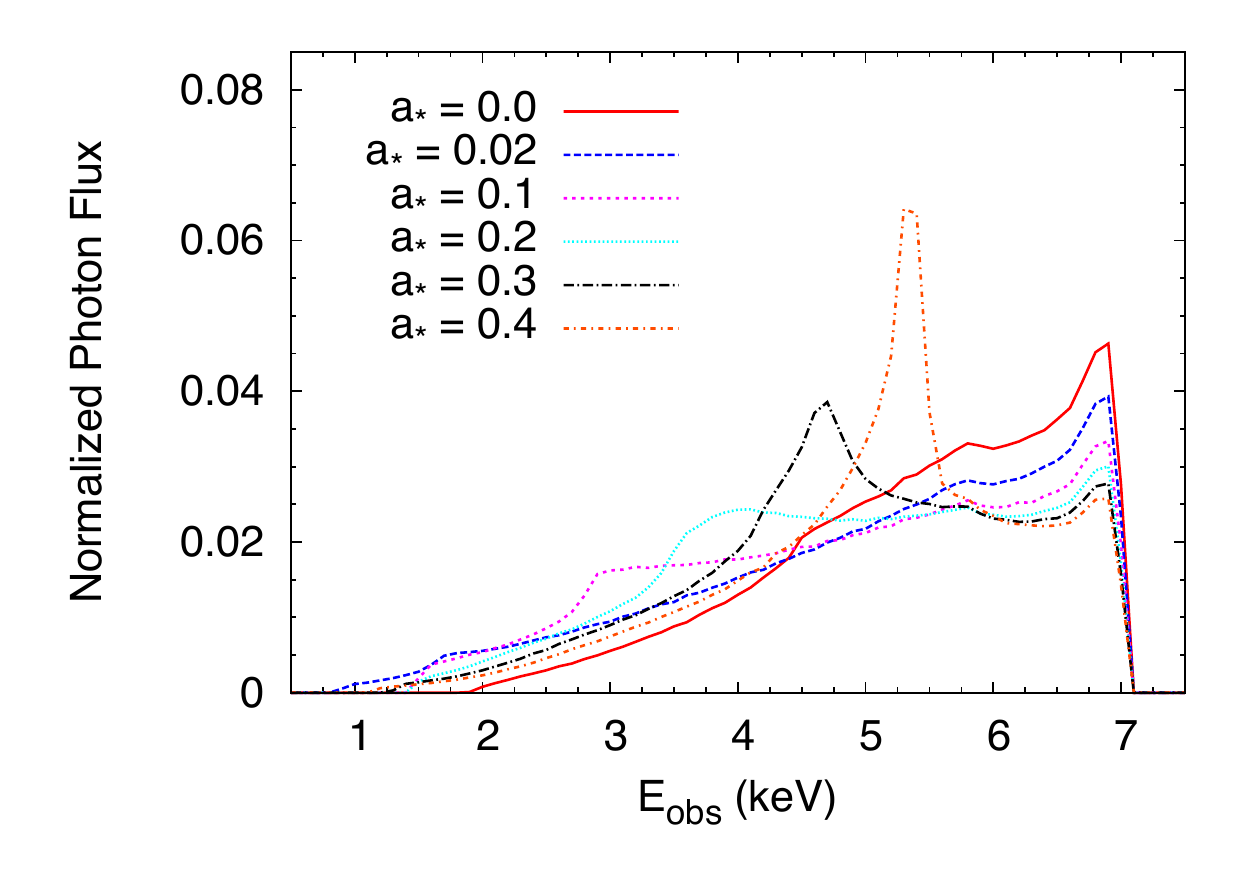}
 \end{center}
\vspace{-0.7cm}
\caption{Broad K$\alpha$ iron line in wormhole backgrounds with $\gamma = 2$ 
(left panel) and $\gamma = 1/2$ (right panel) for different values of the spin $a_*$. 
The astrophysical parameters are: viewing angle $i = 45^\circ$, intensity profile 
with index $\alpha = -3$, and emissivity region with inner radius
$r_{\rm in} = r_{\rm ISCO}$ and outer radius $r_{\rm out} = r_{\rm ISCO} + 100 \, r_0$.
See the text for details.}
\label{f2}
\end{figure*}

\section{Profile of the K$\alpha$ iron line}

The X-ray spectrum of both stellar-mass and supermassive BH candidates 
is usually characterized by the presence of a power-law component. This 
feature is commonly interpreted as the inverse Compton scattering of thermal 
photons by electrons in a hot corona above the accretion disk. The geometry 
of the corona is not known and several models have been proposed. Such 
a ``primary component'' irradiates also the accretion disk, producing a ``reflection 
component'' in the X-ray spectrum. The illumination of the cold disk by the 
primary component also produces spectral lines by fluorescence. The 
strongest line is the K$\alpha$ iron line at 6.4 keV. Especially for some
sources, this line is extraordinarily stable, in spite of a substantial variability 
of the continuum. This fact suggests that its shape is determined by the 
geometry of the spacetime around the compact object.

The profile of the K$\alpha$ iron line depends on the background 
metric, the geometry of the emitting region, the disk emissivity, and the disk's 
inclination angle with respect to the line of sight of the distant observer.
In the Kerr spacetime, the only relevant parameter of the background geometry
is the spin $a_* = J/M^2$, while $M$ sets the length of the system, without affecting 
the shape of the line. In those sources for which there is indication that the line 
is mainly emitted close to the compact object, the emission region may be 
thought to range from the radius of the innermost stable circular orbit (ISCO), 
$r_{\rm in} = r_{\rm ISCO}$, to some outer radius $r_{\rm out}$. However, even 
more complicated geometries are possible. In principle, the disk emissivity 
could be theoretically calculated. In practice, that is not feasible at present. The 
simplest choice is an intensity profile $I_{\rm e} \propto r^{\alpha}$ with index
$\alpha < 0$ to be determined during the fitting procedure. The fourth 
parameter is the inclination of the disk with respect to the line of sight of the 
distant observer, $i$. The dependence of the line profile on $a_*$,
$i$, $\alpha$, and $r_{\rm out}$ in the Kerr background has been analyzed 
in detail by many authors, starting with Ref.~\cite{ka1}. The case of deviations
from the Kerr geometry is discussed in Ref.~\cite{iron}.

Roughly speaking, the calculation of the profile of the K$\alpha$ iron
line goes as follows. We want to compute the photon flux number density 
measured by a distant observer, which is given by
\be
N_{E_{\rm obs}} &=& \frac{1}{E_{\rm obs}} 
\int I_{\rm obs}(E_{\rm obs}) d \Omega_{\rm obs} = \nonumber\\ 
&=& \frac{1}{E_{\rm obs}} \int g^3 I_{\rm e}(E_{\rm e}) 
d \Omega_{\rm obs} \, .
\ee
$I_{\rm obs}$ and $E_{\rm obs}$ are, respectively, the 
specific intensity of the radiation and the photon energy as measured by 
the distant observer, $d \Omega_{\rm obs}$ is the element of the solid 
angle subtended by the image of the disk on the observer's sky,
$I_{\rm e}$ and $E_{\rm e}$ are, respectively, the local specific intensity
of the radiation and the photon energy in the rest frame of the
emitter, and $g = E_{\rm obs}/E_{\rm e}$ is the redshift factor.
$I_{\rm obs} = g^3 I_{\rm e}$ follows from the Liouville's theorem.
As the K$\alpha$ iron line is intrinsically narrow in frequency, we 
can assume that the disk emission is monochromatic (the rest frame energy 
is $E_{\rm{K}\alpha} = 6.4$~keV) and isotropic with a power-law radial 
profile:
\be
I_{\rm e}(E_{\rm e}) \propto \delta (E_{\rm e} - E_{\rm{K}\alpha}) r^{\alpha} \, .
\ee
Doppler boosting, gravitational redshift, and frame dragging are encoded 
in the calculation of $g$, while the light bending enters in the integration.
More details can be found in Ref.~\cite{iron}.

The purpose of this paper is to study how the K$\alpha$ iron line observed 
in the X-ray spectrum of some supermassive BH candidates in galactic
nuclei can test the wormhole nature of these objects. A large number of
wormhole spacetimes have been proposed in the literature~\cite{wh}.
Here, we restrict our attention to traversable wormholes with the line element 
given by~\cite{th,teo}
\begin{widetext}
\be\label{eq-line}
ds^2 = -e^{2\Phi} dt^2 + \frac{dr^2}{1 - b} + r^2 
\left[d\theta^2 + \sin^2\theta \left(d\phi - \omega dt\right)^2 \right] \, ,
\ee
\end{widetext}
where $\Phi$ and $b$ are, respectively, the redshift and the shape
function\footnote{The reader should note that the shape function is 
sometimes defined in a different way in the literature of wormholes.}, 
which, in the general case, may depend on both $r$ and $\theta$.
$\omega$ determines the wormhole angular momentum. In what follows,
I will assume $\Phi = - r_0/r$ and $\omega = 2 J/r^3$, where $r_0$ 
is the throat radius and sets the scales of the system, just like the 
gravitational radius $r_g=M$ does the same for the Kerr background. 
$J$ is the wormhole spin angular momentum. The shape function 
considered in this work has the following form
\be\label{eq-b}
b = \left(\frac{r_0}{r}\right)^\gamma \, , 
\ee
where $\gamma$ is a constant.

The profile of the K$\alpha$ iron line produced in the accretion disk around 
wormholes with $\gamma = 1$ and different values of $a_*=J/r_0^2$ is reported in 
the left panel of Fig.~\ref{f1}. The choice of the value of $\gamma$, and more in
general of the form of $g_{rr}$, changes only the calculation of the photon
trajectories, and therefore the effect of light bending, without altering the
redshift factor $g$. The astrophysical parameters are $i = 45^\circ$, 
$\alpha=-3$, and $r_{\rm out} = r_{\rm ISCO} + 100 \, r_0$. For the reader 
familiar with the profile of the K$\alpha$ iron line produced around 
Kerr BHs, it is straightforward to realize that the line of non-rotating and very
slow-rotating wormholes looks similar to the one of mid- or fast-rotating 
Kerr BHs (see e.g. Fig.~1 in Ref.~\cite{iron}). The line of wormholes with 
slightly higher spin has instead a peculiar low energy bump/peak, 
absent in the Kerr case. Such a peak moves to higher energies as
the wormhole spin increases. The basic properties of the lines in these
wormhole spacetimes and the differences with the Kerr ones can be
understood in terms of ISCO radius and angular velocity of equatorial
circular orbits.

As already discussed in~\cite{th}, in these spacetimes 
$r_{\rm ISCO} = 2 \, r_0$ for $a_* = 0$ and decreases regularly to 
$r_{\rm ISCO} = 1.29 \, r_0$ for $a_* = 0.016693$. For higher values
of the spin, equatorial circular orbits are always stable, and therefore the
inner radius of the disk is at $r_0$. This behavior should be compared with 
the one around a Kerr BH, noting that $r_0$ plays the role of $M$.
In Kerr, $r_{\rm ISCO} = 6 \, M$ for $a_* = 0$ and decreases regularly
as the spin parameter increases, till $r_{\rm ISCO} = M$ for $a_* = 1$.
For instance, $r_{\rm ISCO} = 2 \, M$ when $a_* \approx 0.943$.
This explains the low energy tail in the wormhole line: even for non-rotating
or very slow-rotating wormholes, the inner edge of the disk is at very
small radii and therefore the line is affected by a strong gravitational redshift.

The low energy bump in the wormhole line, which becomes a pronounced
peak when $a_* \gtrsim 0.2$, is also generated at very small radii. This can 
be easily checked by calculating the contribution of the photons emitted at
small radii, see Fig.~\ref{f1b}. The high Doppler boosting is a consequence 
of the quick increase of the Keplerian angular velocity $\Omega$. The right 
panel of Fig.~\ref{f1} shows $\Omega$ for a Kerr BH with spin $a_*=1$ and 
for some traversable wormholes. For low and mid values of $a_*$, at radii 
$r > 2 \, r_0$ the angular velocity in wormhole spacetimes is quite independent 
of the wormhole spin and it is lower than the Kerr one. In such spacetimes, 
the wormhole spin plays an important role when $r < 2 \, r_0$, and $\Omega$ 
increases quickly as the radius decreases. The angular velocity can exceed 
the maximum angular velocity of the Kerr background $\Omega_{\rm Kerr, \, max} 
= 1/(2M)$. The peak moves to higher energy as the wormhole spin increases, 
as the angular velocity at small radii also increases. If the value of the wormhole
spin is high (e.g. $a_*=1$), the peak produced by the Doppler boosting at
small radii may be confused with the one of a Kerr BH produced at larger radii
(see the right panel in Fig.~\ref{f1b}). However, in this case the wormhole
line presents also a high energy tails, completely absent when the compact
object is a Kerr BH.

\begin{figure*}
\begin{center}
\includegraphics[type=pdf,ext=.pdf,read=.pdf,width=7.5cm]{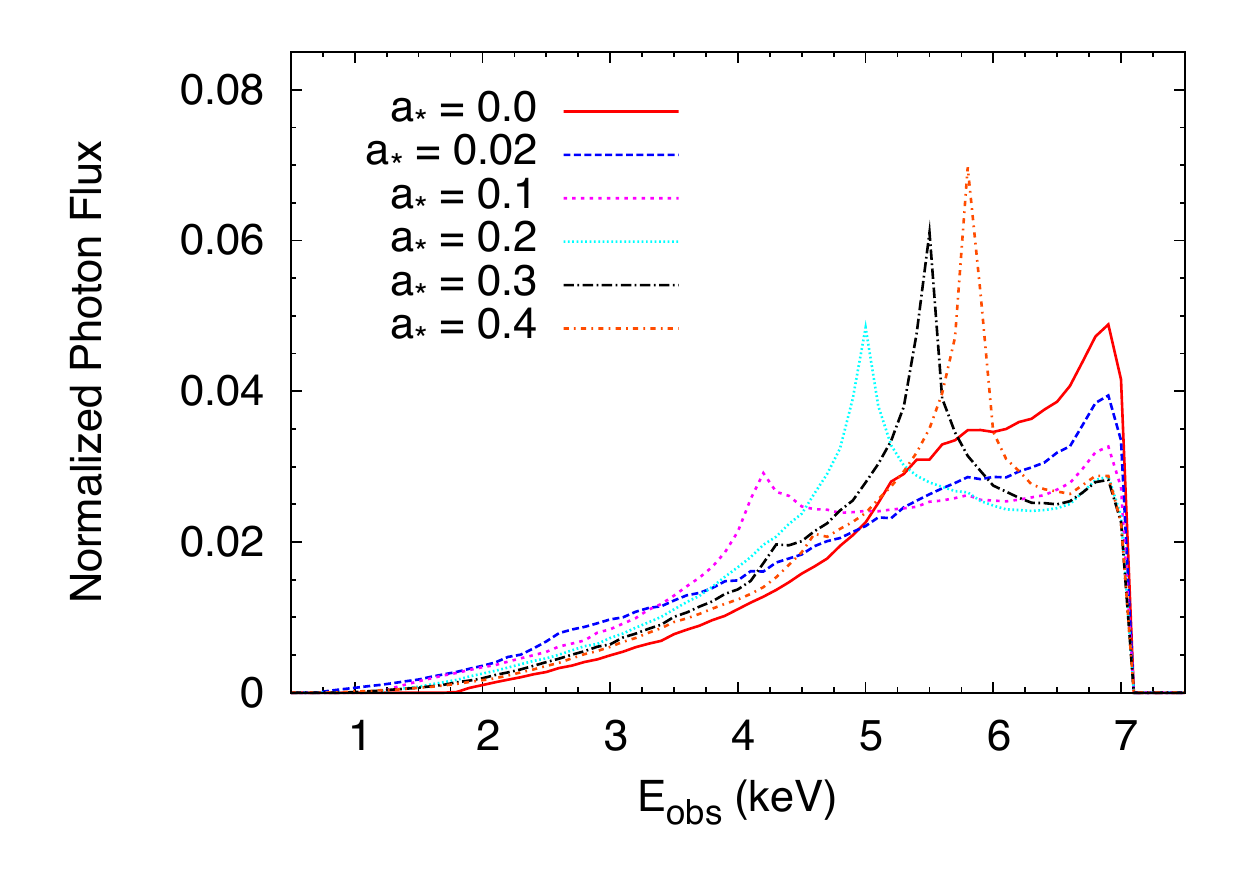}
\includegraphics[type=pdf,ext=.pdf,read=.pdf,width=7.5cm]{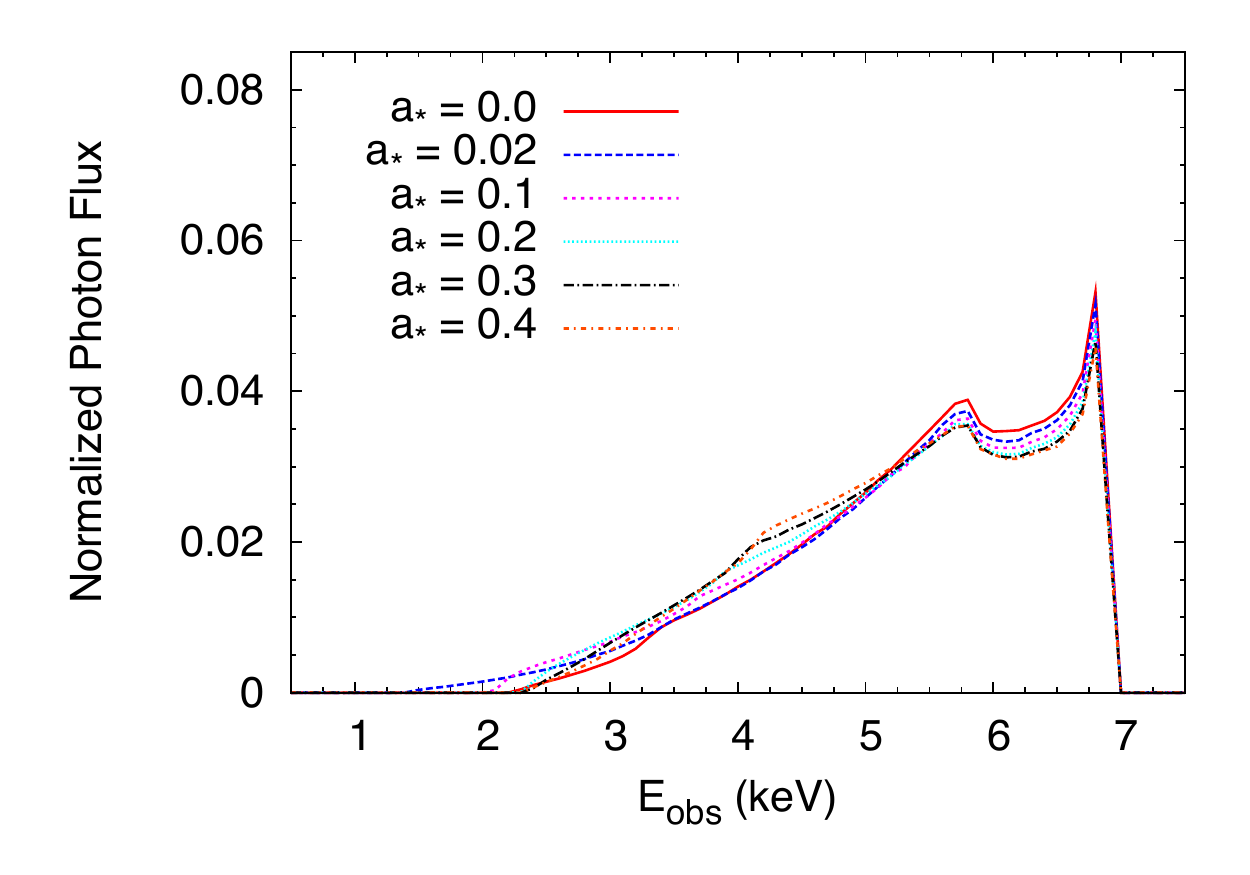}
 \end{center}
\vspace{-0.7cm}
\caption{Broad K$\alpha$ iron line in wormhole backgrounds with $\gamma = 100$ 
(left panel) and $\gamma = 1/100$ (right panel) for different values of the spin $a_*$. 
The astrophysical parameters are: viewing angle $i = 45^\circ$, intensity profile 
with index $\alpha = -3$, and emissivity region with inner radius
$r_{\rm in} = r_{\rm ISCO}$ and outer radius $r_{\rm out} = r_{\rm ISCO} + 100 \, r_0$.
See the text for details.}
\label{f2b}
\end{figure*}

If we change the value of the parameter $\gamma$ in Eq.~(\ref{eq-b}), we
only affect the propagation of the photons from the disk to the observer,
without altering the redshift factor, which does not depend on $g_{rr}$.
Fig.~\ref{f2} shows the expected K$\alpha$ iron line for traversable
wormholes with $\gamma=2$ (left panel) and $1/2$ (right panel). The
qualitative properties of the line are the same of the case $\gamma = 1$.
The line produced around non-rotating or very slow-rotating wormholes 
has the low energy tail similar to one expected from mid- or
fast-rotating Kerr BHs. For slightly larger values of $a_*$, the line presents
the low energy peak due to the gravitational redshift and Doppler blueshift 
at small radii. The limiting cases of very large and very small $\gamma$
can be understood from the panels in Fig.~\ref{f2b}. The left panel shows the
case $\gamma = 100$ and it is qualitatively very similar to the previous ones.
Indeed, here the propagation of the photon is altered only in a 
very small region close to the wormhole throat ($r_0/r$ is always smaller 
than 1). The right panel in Fig.~\ref{f2b} shows the opposite case, of very small 
$\gamma$, specifically with $\gamma = 1/100$. Now the propagation of 
photons is altered even at large radii and the final effect is to significantly
change the low energy peak, which becomes a small bump around 4~keV,
quite independently of the value of the spin (the two peaks at $\sim 6$~keV 
and $\sim 7$~keV are instead the result, respectively, of the Doppler
redshift and blueshift at larger radii).

\begin{table*}
\begin{center}
\begin{tabular}{c c c c c}
\hline
\hline
AGN &  \hspace{.5cm} & $a_*$ &  \hspace{.5cm} & References \\
\hline
\hline
MGC-6-30-15 && $> 0.98$ && \cite{mcg} \\
Fairall~9 && $0.65\pm0.05$ && \cite{F9,patrick} \\
SWIFT~J2127.4+5654 && $0.6\pm0.2$ && \cite{2127} \\
1H~0707-495 && $> 0.98$ && \cite{0707} \\
Mrk~79 && $0.7\pm0.1$ && \cite{79} \\
NGC~3783 && $> 0.98$ && \cite{3783} \\
Mrk~335 && $0.70\pm0.12$ && \cite{patrick} \\
NGC~7469 && $0.69\pm0.09$ && \cite{patrick} \\
\hline
\hline
\end{tabular}
\end{center}
\vspace{-0.2cm}
\caption{Current measurements of the spin parameter of supermassive BH
candidates with the analysis of the K$\alpha$ iron line.}
\label{tab}
\end{table*}

\begin{figure*}
\begin{center}
\vspace{-1.0cm}
\includegraphics[type=pdf,ext=.pdf,read=.pdf,width=7.5cm]{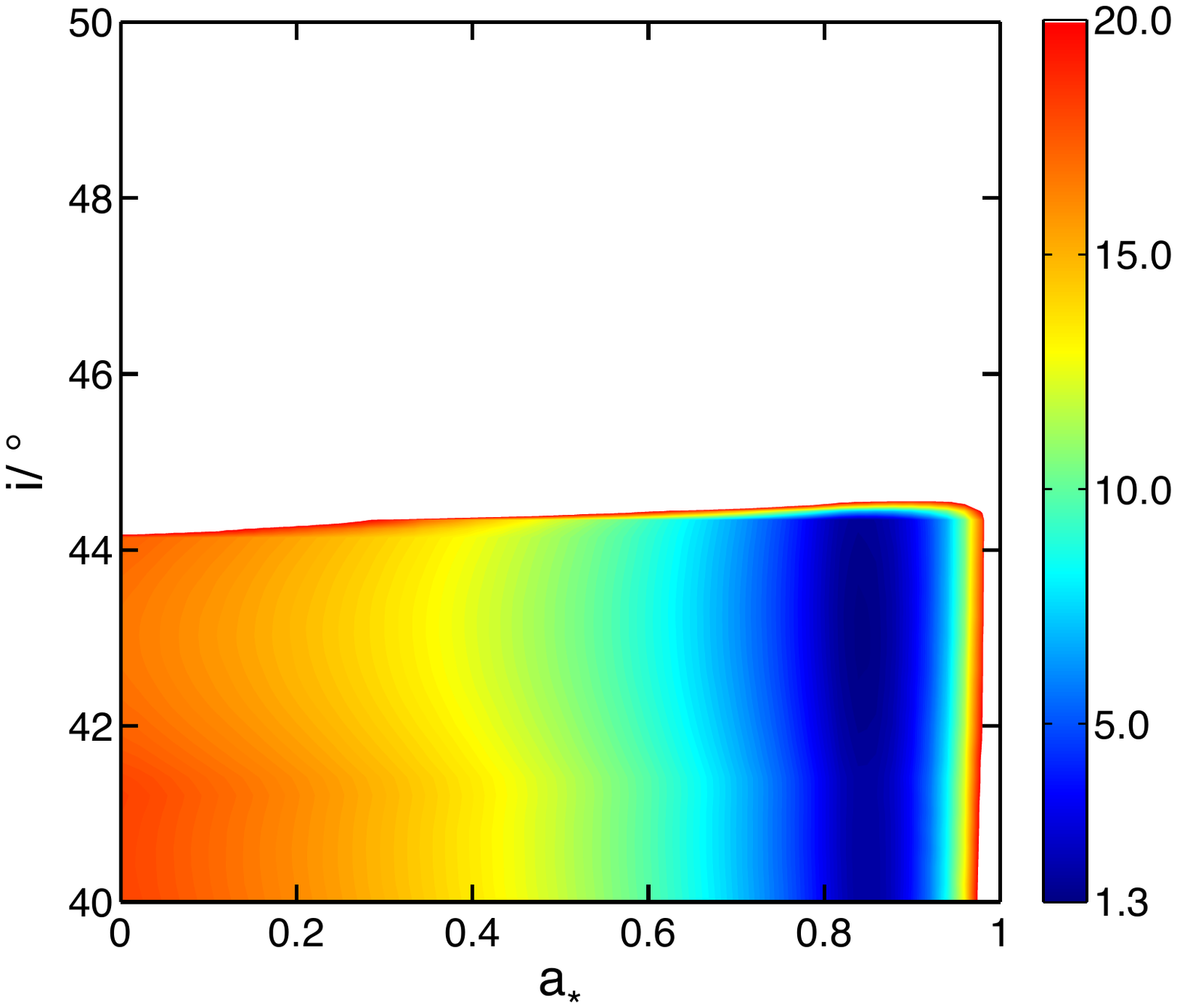}
\includegraphics[type=pdf,ext=.pdf,read=.pdf,width=7.5cm]{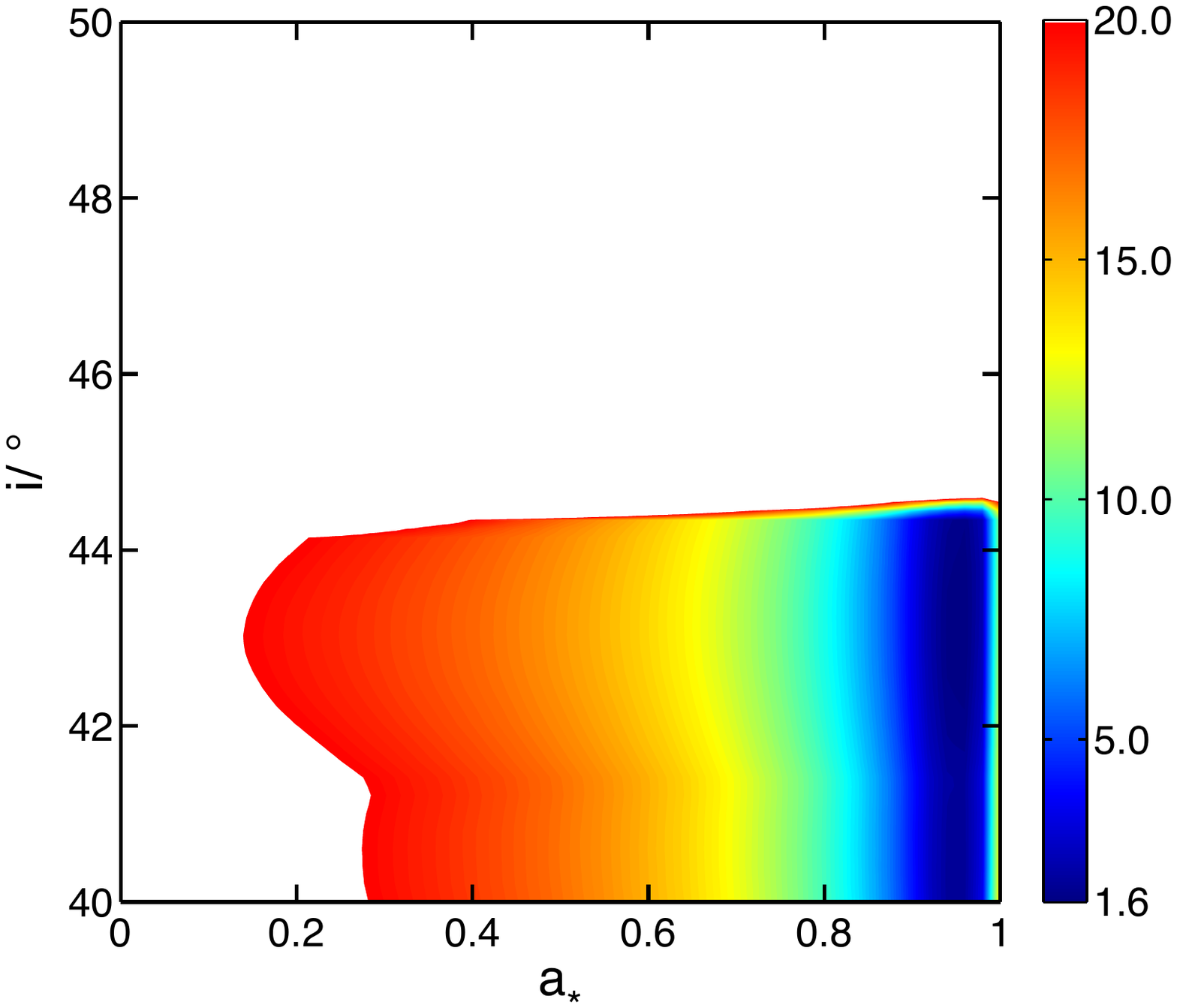} \\
\vspace{-4.0cm}
\includegraphics[type=pdf,ext=.pdf,read=.pdf,width=7.5cm]{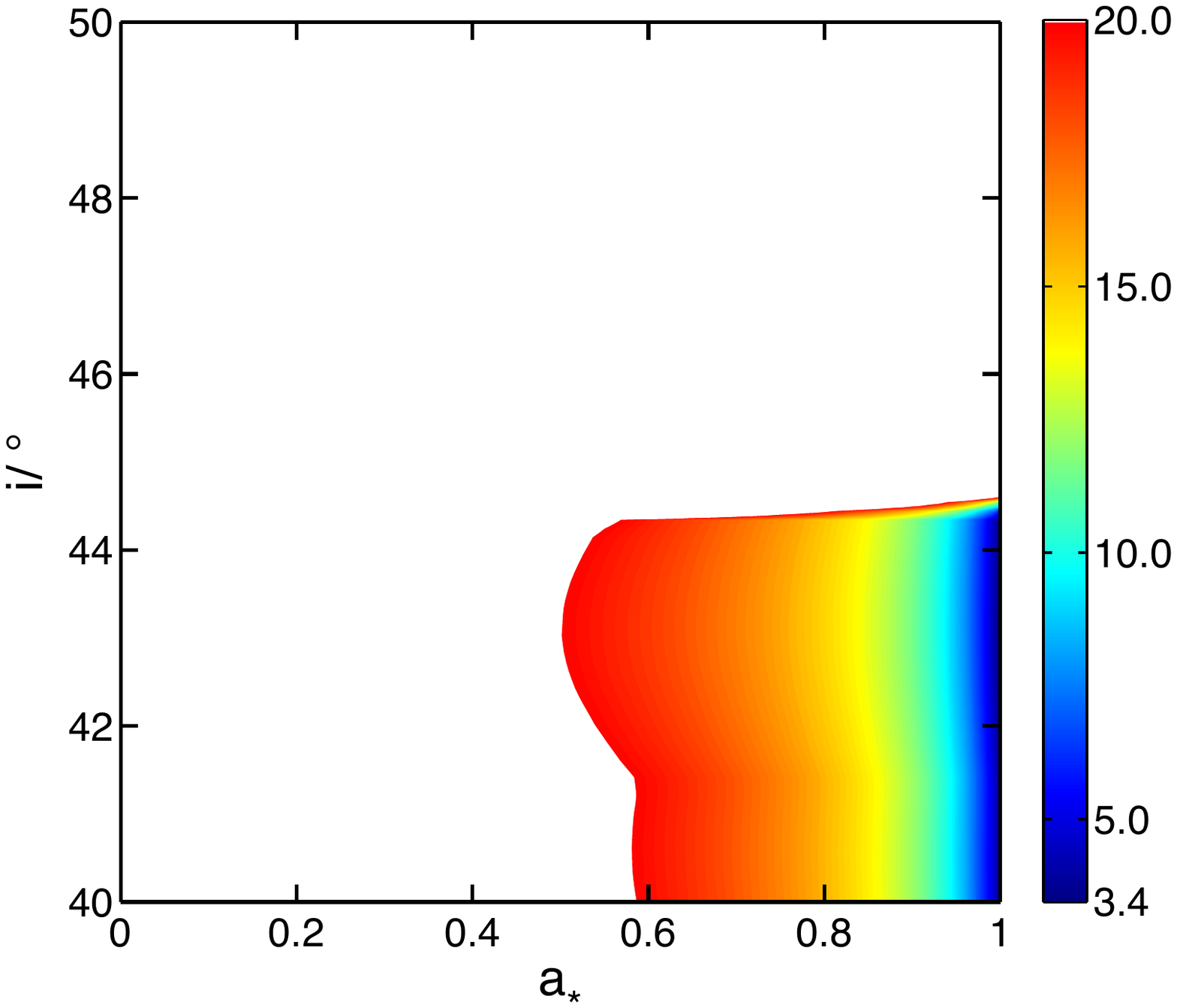}
\includegraphics[type=pdf,ext=.pdf,read=.pdf,width=7.5cm]{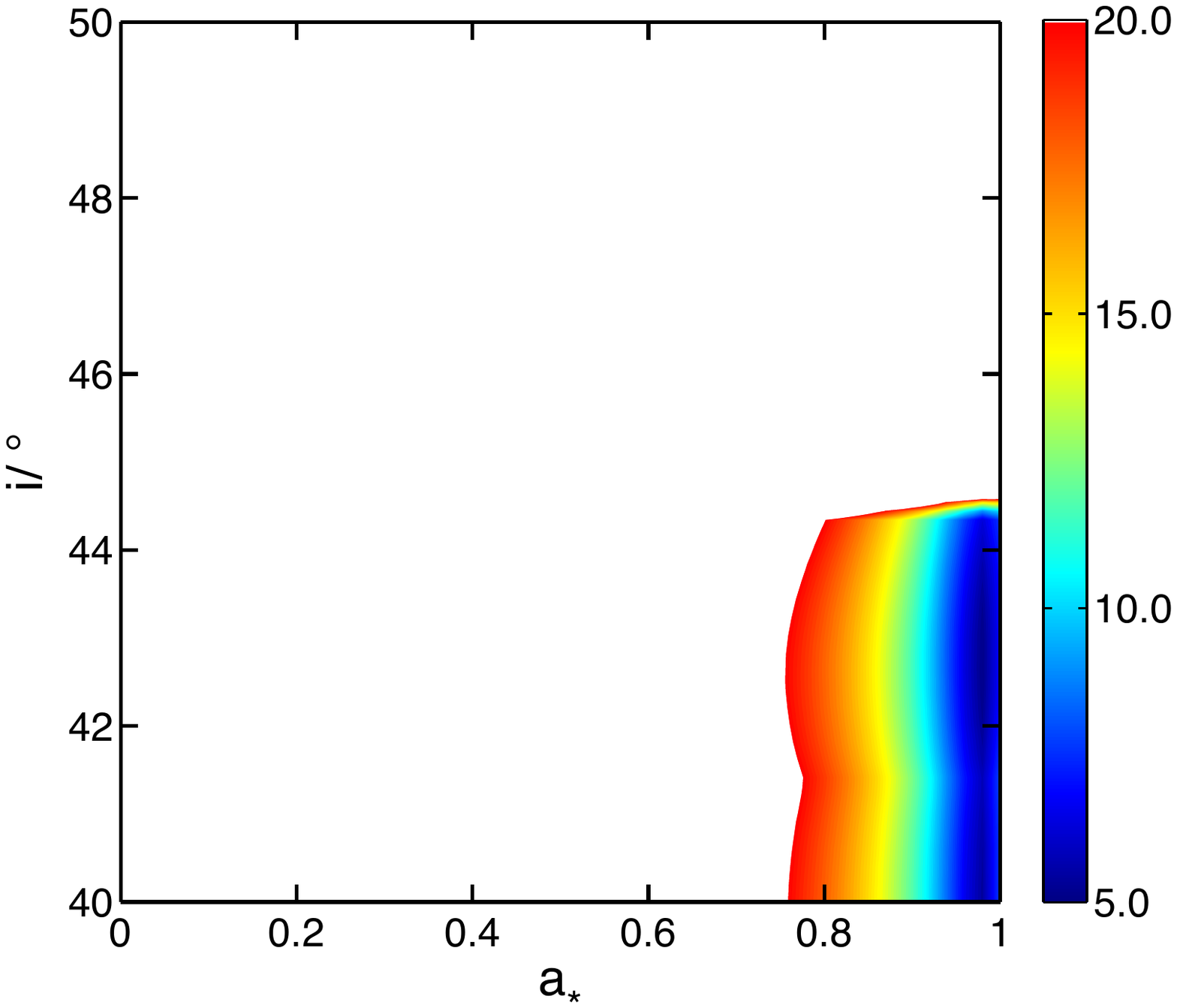}
 \end{center}
 \vspace{-2.5cm}
\caption{Reduced $\chi^2$ from the comparison of the profile of the
K$\alpha$ iron line produced in a Kerr spacetime with spin parameter
$a_*$ and observed with a viewing angle $i$ and the line generated
in a wormhole background with $\gamma = 1$, spin parameter 
$\tilde{a}_* = 0$ (top left panel), 0.015 (top right panel), 0.02 (bottom 
left panel), and 0.3 (bottom right panel), and observed with a viewing 
angle $\tilde{i}=45^\circ$. The intensity profile has index $\alpha = 
\tilde{\alpha} = -3$ and the emissivity region has an inner radius at the ISCO 
and $r_{\rm out} - r_{\rm ISCO} = \tilde{r}_{\rm out} - \tilde{r}_{\rm ISCO} 
= 100 \, r_0$. See the text for details.}
\label{f3}
\end{figure*}

\begin{figure*}
\begin{center}
\vspace{-1.0cm}
\includegraphics[type=pdf,ext=.pdf,read=.pdf,width=7.5cm]{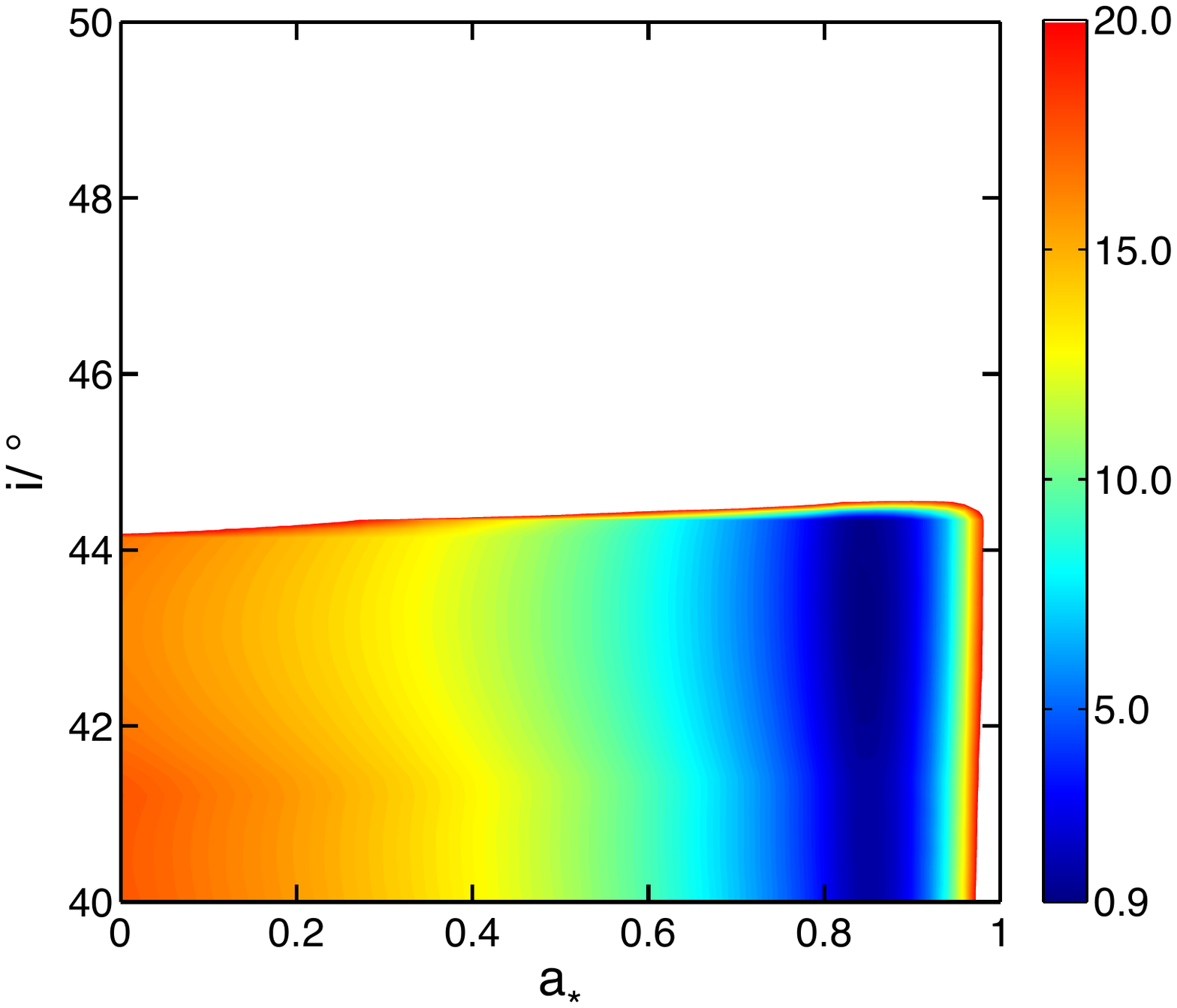}
\includegraphics[type=pdf,ext=.pdf,read=.pdf,width=7.5cm]{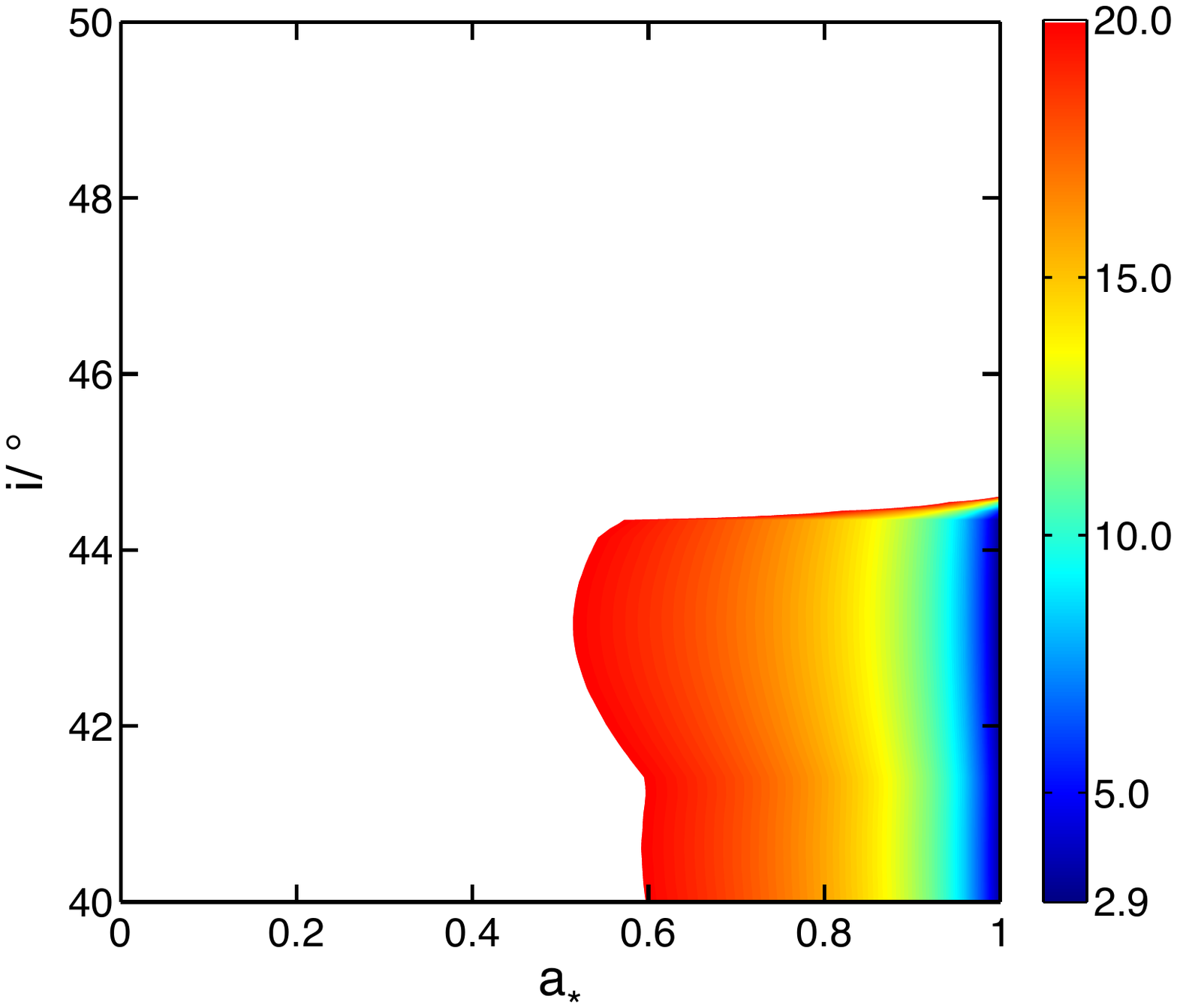}
 \end{center}
 \vspace{-2.5cm}
\caption{As in Fig.~\ref{f3}, for the case $\gamma = 2$. The spin parameter
is $\tilde{a}_* = 0$ (left panel) and 0.02 (right panel).}
\label{f4}
\end{figure*}

\begin{figure*}
\begin{center}
\vspace{-1.0cm}
\includegraphics[type=pdf,ext=.pdf,read=.pdf,width=7.5cm]{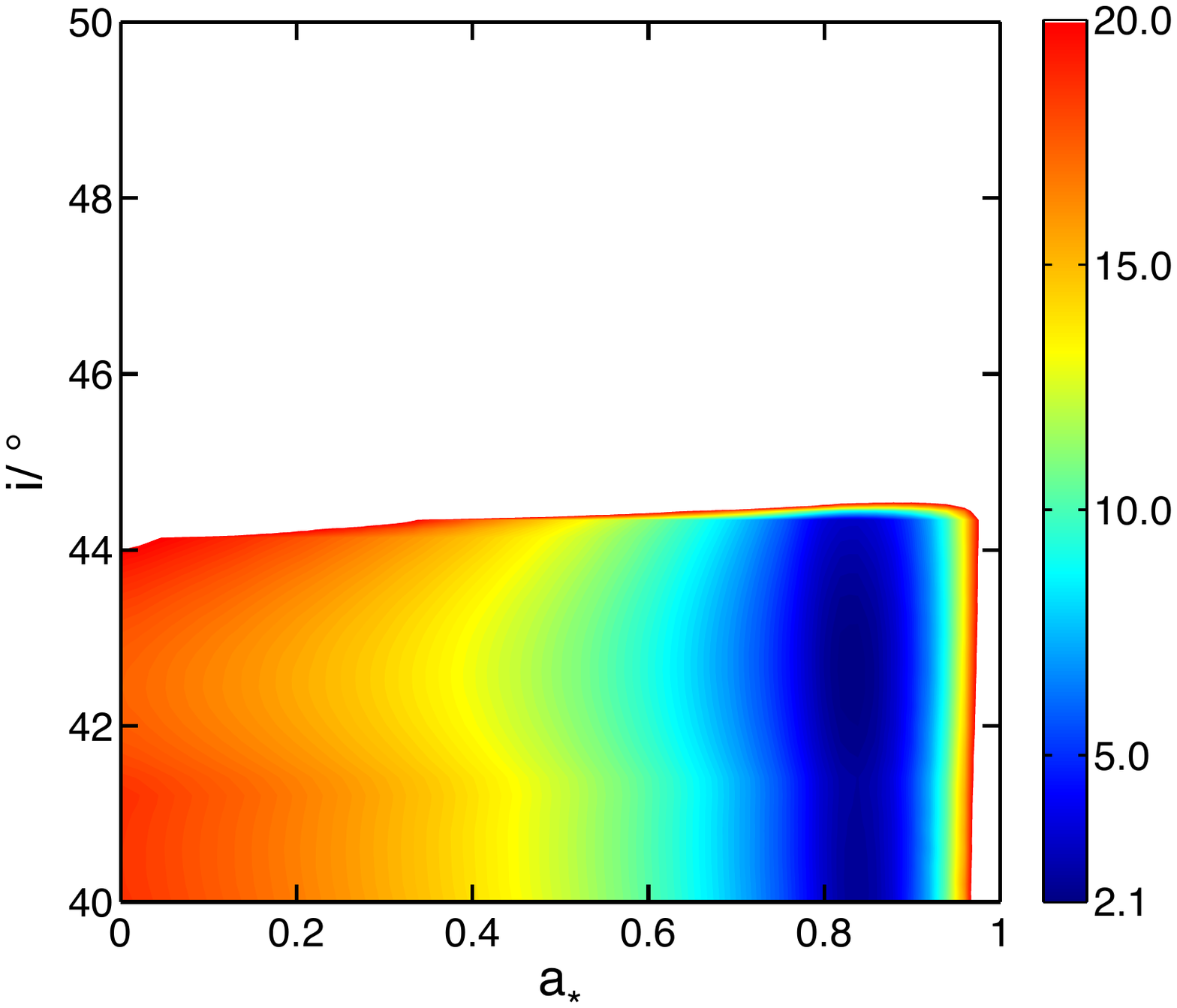}
\includegraphics[type=pdf,ext=.pdf,read=.pdf,width=7.5cm]{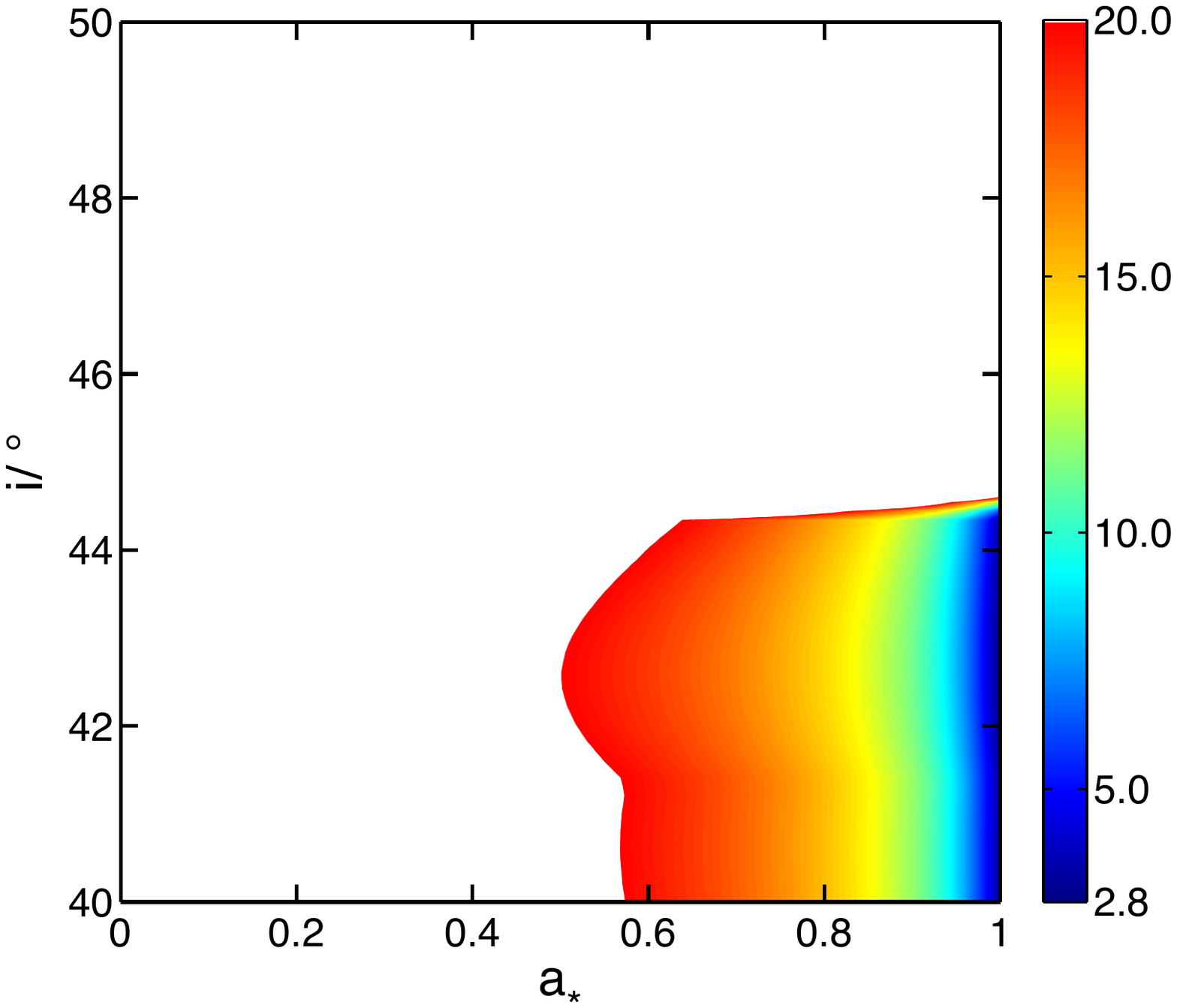}
 \end{center}
 \vspace{-2.5cm}
\caption{As in Fig.~\ref{f3}, for the case $\gamma = 1/2$. The spin parameter
is $\tilde{a}_* = 0$ (left panel) and 0.02 (right panel).}
\label{f5}
\end{figure*}

\begin{figure*}
\begin{center}
\vspace{-1.0cm}
\includegraphics[type=pdf,ext=.pdf,read=.pdf,width=7.5cm]{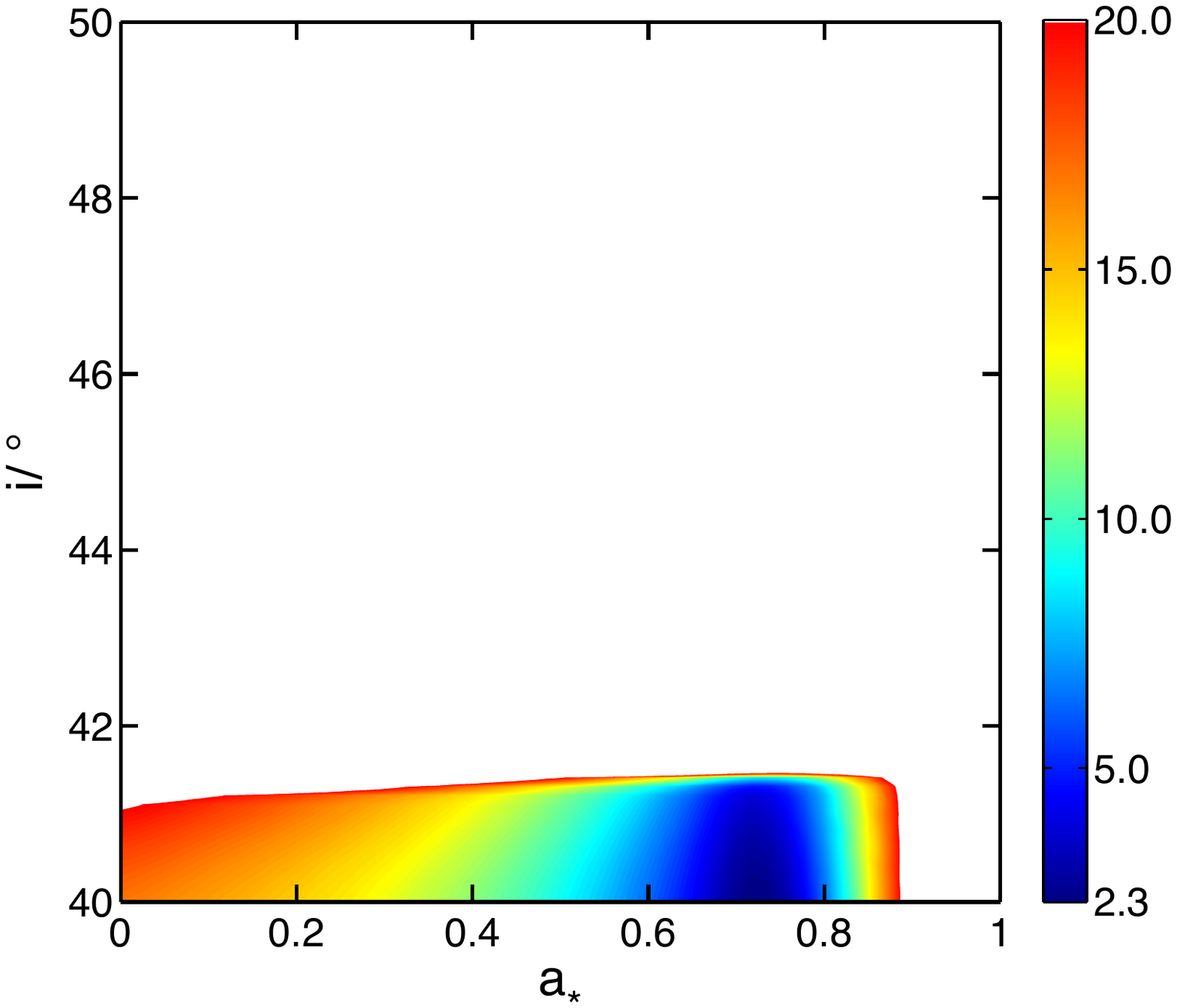}
\includegraphics[type=pdf,ext=.pdf,read=.pdf,width=7.5cm]{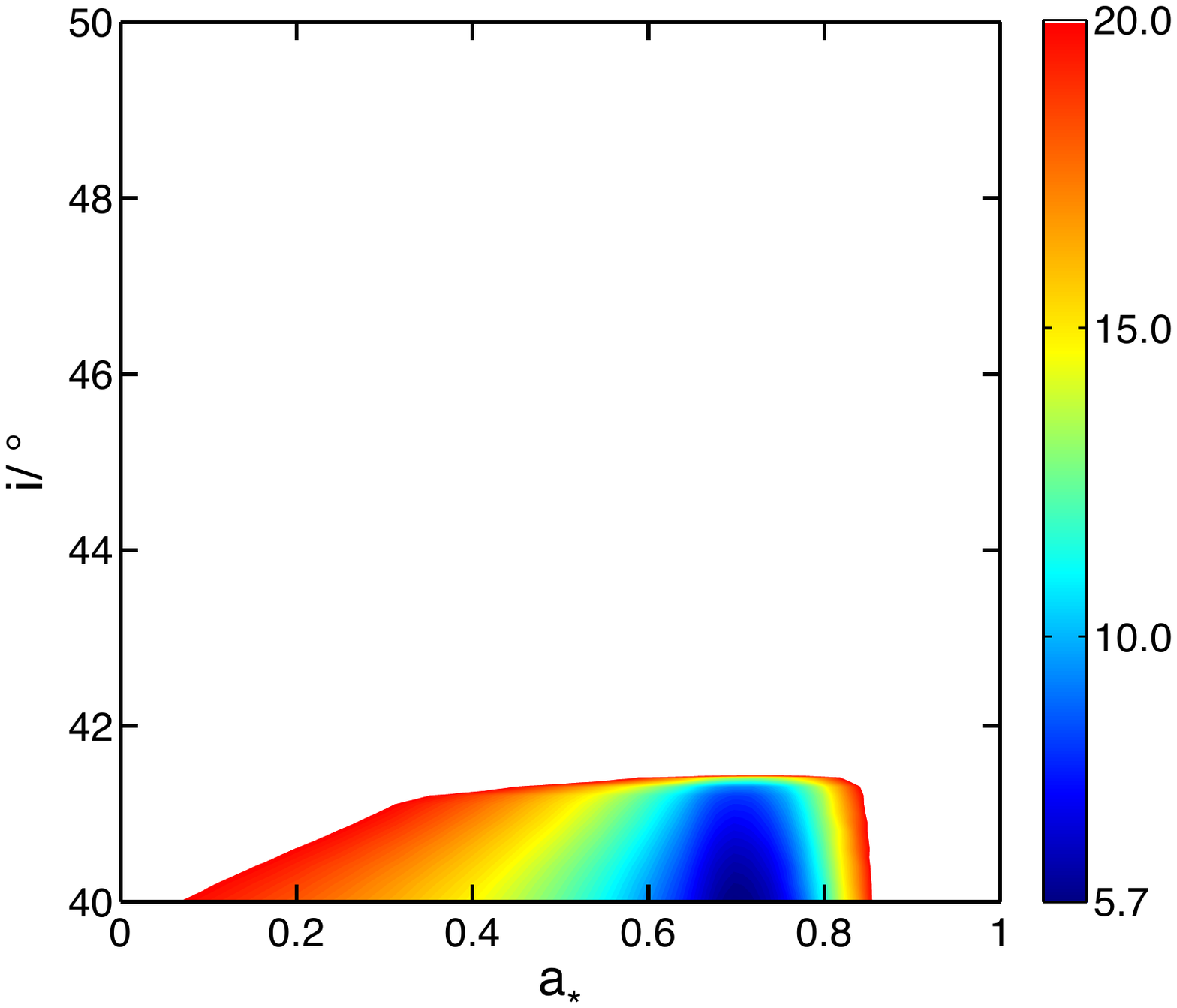}
 \end{center}
 \vspace{-2.5cm}
\caption{As in Fig.~\ref{f3}, for the case $\gamma = 1/100$. The spin parameter
is $\tilde{a}_* = 0$ (left panel) and 0.4 (right panel).}
\label{f6}
\end{figure*}

\begin{table*}
\begin{center}
\begin{tabular}{c c c c | c c c c c c}
\hline
\hline
$\gamma$ &  \hspace{.5cm} & $\tilde{a}_*$ &  \hspace{.5cm} &  \hspace{.5cm} & $\chi^2_{\rm red, \, min}$ &  \hspace{.5cm} & $a_*$ &  \hspace{.5cm} & $i$ \\
\hline
\hline
1 && 0 &&& 1.31 && 0.84 && 43.2$^\circ$ \\
1 && 0.015 &&& 1.62 && 0.96 && 43.4$^\circ$ \\
1 && 0.02 &&& 3.43 && 1 && 43.2$^\circ$ \\
1 && 0.3 &&& 5.69 && 0.98 && 42.6$^\circ$ \\
2 && 0 &&& 0.90 && 0.84 && 43.2$^\circ$ \\
2 && 0.02 &&& 2.87 && 1 && 43.2$^\circ$ \\
1/2 && 0 &&& 2.06 && 0.84 && 42.4$^\circ$ \\
1/2 && 0.02 &&& 2.79 && 1 && 42.8$^\circ$ \\
1/100 && 0 &&& 2.27 && 0.72 && 40.0$^\circ$ \\
1/100 && 0.4 &&& 5.64 && 0.70 && 40.0$^\circ$ \\
\hline
\hline
\end{tabular}
\end{center}
\vspace{-0.2cm}
\caption{Minimum of the reduced $\chi^2$ and the corresponding Kerr
values of $a_*$ and $i$ for the lines produced by wormholes with $\gamma$
and $\tilde{a}_*$ given in the first and second column, and 
$\tilde{i} = 45^\circ$. See the text for more details.}
\label{tab2}
\end{table*}

\section{Discussion}

The analysis of the K$\alpha$ iron line is commonly used to estimate the spin
parameter of a BH candidate, under the assumption that the 
geometry of the spacetime around the object is described by the Kerr solution. 
The measurements reported in the literature of the supermassive BH 
candidates are shown in Tab.~\ref{tab}. Roughly speaking, there are three 
objects that seem to be very fast-rotating Kerr BHs ($a_*>0.98$), and five 
objects that are consistent with Kerr BHs with a mid value of the spin 
parameter ($a_* \sim 0.6 - 0.8$).

What happens if the supermassive BH candidates are wormholes but we
assume they are Kerr BHs and we try to estimate their spin parameter? Can
a wormhole be confused with a Kerr BH of different spin? A qualitative
answer to these questions has been already provided in the discussion 
in the previous section, but here we want to be more quantitative. We can
compare the expected K$\alpha$ iron line from a wormhole spacetime with
the ones from Kerr BHs. We use the same approach of Ref.~\cite{iron}
and we define the reduced $\chi^2$ as
\begin{widetext}
\be\label{eq-chi2}
\chi^2_{\rm red} (a_*, i, \alpha, r_{\rm out})
= \frac{\chi^2}{n} =
\frac{1}{n} \sum_{i = 1}^{n} \frac{\left[N_i^{\rm Kerr} 
(a_*, i, \alpha, r_{\rm out}) - N_i^{\rm WH}
(\tilde{a}_*, \tilde{i}, \tilde{\alpha}, \tilde{r}_{\rm out}) 
\right]^2}{\sigma^2_i} \, ,
\ee 
where the summation is performed over $n$ sampling energies $E_i$ and
$N_i^{\rm Kerr}$ and $N_i^{\rm WH}$ are the normalized photon fluxes in the 
energy bin $[E_i,E_i+\Delta E]$ respectively for the Kerr and the wormhole metric. 
Here the error $\sigma_i$ is assumed to be 15\% the normalized photon flux
$N_i^{\rm WH}$,
\be
\sigma_i = 0.15 \, N_i^{\rm WH} \, ,
\ee
which is roughly the accuracy of current observations in the best situations.
For the calculation of $N_i^{\rm Kerr}$, we use the general form of the Kerr
metric
\be\label{eq-kerr}
ds^2 = \left(1 - \frac{2 m r}{\Sigma}\right) dt^2 
+ \frac{4 a m r \sin^2\theta}{\Sigma} dt d\phi
- \frac{\Sigma}{\Delta} dr^2 - \Sigma d\theta^2
- \sin^2 \theta \left(r^2 + a^2 + \frac{2 a^2 m r 
\sin^2\theta}{\Sigma} \right) d\phi^2 \, ,
\ee
\end{widetext}
where $a = J/M$, $\Sigma = r^2 + a^2 \cos^2\theta$, and $\Delta = r^2 - 2 m r + a^2$, 
so both Eqs.~(\ref{eq-line}) and (\ref{eq-kerr}) are valid for arbitrary values of 
the spin parameter (but for the Kerr metric it makes sense to consider only the case 
$|a_*| \le 1$).

Let us start considering a wormhole background with $\gamma = 1$. In this
simplified analysis, we assume $\alpha = \tilde{\alpha}$ and $r_{\rm out} =
\tilde{r}_{\rm out}$. The
reduced $\chi^2$ for wormholes with spin parameter $\tilde{a}_* = 0$, 0.015,
0.02, and 0.3 and viewing angle $\tilde{i} = 45^\circ$ is shown in Fig.~\ref{f3}.
The values of $a_*$ and $i$ at the minimum of the reduced $\chi^2$ are
reported in Tab.~\ref{tab2}. The minimum of $\chi^2_{\rm red}$ is at a
viewing angle $i$ slightly lower than $\tilde{i}$ because the wormhole
angular frequency at large radii is lower than the Kerr one and the 
high-energy peak of the line is produced by the Doppler blueshift at
relatively large radii. For non-rotating or very slow-rotating wormholes
($\tilde{a}_* = 0$ and 0.015), the minimum of $\chi^2_{\rm red}$ is around
1, which means that these lines may well fit the ones expected for a Kerr
BH, providing, however, a completely wrong estimate of the spin $a_*$.
Indeed, $a_*$ is in the range $0.84-1$. Such a value of the spin parameter
is marginally consistent with the current measurements shown in Tab.~\ref{tab}.
For $a_* = 0.02$, the fit is already quite bad, as the minimum of 
$\chi^2_{\rm red}$ is $\sim 3$ (bottom left panel in Fig.~\ref{f3}). As the wormhole 
spin increases, the fit becomes worse and worse. If $\tilde{a}_* = 0.3$, the 
minimum of $\chi^2_{\rm red}$ is $\sim 5$ (bottom right panel in Fig.~\ref{f3}). 
For high values of the wormhole spin, e.g. $a_*=1$, the peak produced by
Doppler boosting at small radii may look like the one of a Kerr BH. However,
as already pointed out in the previous section, there is now a high energy tail
absent in the Kerr line. The fit is therefore bad: for instance, if $a_*=1$,
the minimum of $\chi^2_{\rm red}$ is $\sim 8$. While a rigorous analysis 
would require us to consider real data of specific sources, the fact that
current X-ray data give good fits when the Kerr metric is assumed and that
already for $a_* \gtrsim 0.02$ we find $\chi^2_{\rm red,\, min} > 3$ can be 
used to conclude that rotating wormholes with a moderate value of the spin 
parameter may already be ruled out as candidates to explain the supermassive
objects at the centers of galaxies.

For a different value of $\gamma$, the situation is similar (except for very small
values of $\gamma$), as shown in Fig.~\ref{f4} and \ref{f5}. Non-rotating wormholes 
with higher values of $\gamma$ seem to produce spectra more similar to the 
one expected from Kerr BHs. When $\gamma$ is very small, one finds contour plots
like the ones in Fig.~\ref{f6} (some caution in the interpretation of these plots is
necessary here, as $\chi^2_{\rm red,\, min}$ is for $i < 40^\circ$, but from the
contour plots of the previous cases, it is straightforward to guess the behavior for
lower angles). Despite the quite different line for very low values of $\gamma$,
for non-rotating wormholes we still find marginally acceptable fits, while for 
rotating wormholes the fits are bad.

As a final remark, we can note that the K$\alpha$ iron line analysis can also
be used to test the nature of stellar-mass BH candidates and verify if they
are Kerr BHs or traversable wormholes, even if the possibility of the 
existence of stellar-mass traversable wormholes seems to be less
theoretically motivated. The iron line approach cannot instead be used to
probe the geometry of the spacetime around stellar-type compact objects
like neutron stars, as in this case the inner edge of the accretion disk is
determined by the magnetosphere, rather than by the background metric,
and it is at larger radii.

\section{Summary and conclusions}

It has been proposed that the supermassive objects at the centers of galaxies
may be wormholes formed in the early Universe and connecting either our 
Universe with other Universes, or two different regions of our Universe. The
wormhole paradigm may explain the observations of very massive objects already
at high redshift, as well as the non-observation of thermal radiation from the
putative surface of these objects. In this paper, I have investigated how the
analysis of the profile of the K$\alpha$ iron line can test the wormhole-nature of
the supermassive BH candidates. While observational tests to distinguish
wormholes from Kerr BHs have already been discussed by other authors, the
K$\alpha$ iron line approach has never been considered so far, despite the fact that
it is currently the only available technique to probe the spacetime geometry
around these objects.

The traversable wormholes discussed in this paper have the line element given
in Eq.~(\ref{eq-line}). Their most important feature is that the ISCO radius is
close to the wormhole throat already in the non-rotating case. The radiation
emitted in the inner part of the accretion disk is thus strongly redshifted. This
leads to a low energy tail in the line seen by a distant observer similar to the
one of a Kerr BH with a mid or high value of $a_*$. Already for very moderate 
values of the wormhole spin parameter, the angular velocity of equatorial
circular orbits increases quickly at small radii, producing a low energy peak in 
the observed line. Such a feature is not present in the case of Kerr BHs and
thus represents an important observational signature of rotating wormholes.
However, a similar feature has never been observed in the X-ray spectrum of
BH candidates. A more quantitative analysis of the comparison of the line
expected from a wormhole with the one expected from a Kerr BH confirms
this picture, see Figs.~\ref{f3}, \ref{f4}, \ref{f5}, and \ref{f6}. Non-rotating or very 
slow-rotating wormholes may be confused with mid- or fast-rotating Kerr BHs 
by current observations (but with future more accurate observations we should 
be able to distinguish the two cases). The wormholes with slightly higher 
value of the spin parameter are distinguishable from Kerr BHs by current 
observations and they may already be ruled out. This conclusion is based 
on the fact that current analysis of the K$\alpha$ iron line of some supermassive 
BH candidates produce good fits ($\chi^2_{\rm red,\, min} \sim 1$) when the 
Kerr background is assumed~\cite{mcg,F9,patrick,2127,0707,79,3783}. The 
study presented in this work does not exclude all the rotating wormholes, but 
only the one with the line element given by Eq.~(\ref{eq-line}). For instance, for 
wormholes traversable in only one direction, the outside geometry may be 
the same as the one of Kerr BHs~\cite{niko}.


\begin{acknowledgments}
I thank Zilong Li for help in the preparation of the manuscript.
This work was supported by the Thousand Young Talents 
Program and Fudan University.
\end{acknowledgments}


\end{document}